\documentstyle[emulateapj]{article}
\newcommand{\begit}{\begin{itemize}}
\newcommand{\enit}{\end{itemize}}
\newcommand{\begen}{\begin{enumerate}}
\newcommand{\enen}{\end{enumerate}}

\newcommand       \be           {\begin{equation}}
\newcommand       \ee           {\end{equation}}
\newcommand       \bea          {\begin{eqnarray}}
\newcommand       \eea          {\end{eqnarray}}

\newcommand       \kms		{\,{\rm km \,\, s}^{-1}}
\newcommand       \cm		{\,{\rm cm }}
\newcommand       \pc		{\,{\rm pc }}
\newcommand       \K		{\,{\rm K }}
\newcommand       \yr		{\,{\rm yr }}
\newcommand       \s		{\,{\rm s }}
\newcommand       \kpc		{\,{\rm kpc }}
\newcommand       \erg		{\,{\rm erg }^{-1}}
\newcommand       \ergs		{\,{\rm erg \,\, s}^{-1}}
\newcommand{\beqa}{\begin{eqnarray}} 
\newcommand{\eeqa}{\end{eqnarray}}

\begin{document}

\title{THE IONIZATION STATE OF SODIUM IN GALACTIC WINDS}

\author{Norman Murray\altaffilmark{1,2}, Crystal
  L. Martin\altaffilmark{3,4,5}, Eliot Quataert\altaffilmark{3,4,6}, \& Todd A.~Thompson\altaffilmark{7,8}}

\altaffiltext{1}{Canada Research Chair in Astrophysics}
\altaffiltext{2}{Canadian Institute for Theoretical Astrophysics, 60 St.~George Street, University of Toronto, Toronto,
ON M5S 3H8, Canada; murray@cita.utoronto.ca}
\altaffiltext{3}{Packard Fellow}
\altaffiltext{4}{Alfred P. Sloan Research Fellow}
\altaffiltext{5}{University of California, Santa Barbara, Department
  of Physics, Santa Barbara, CA 93106; cmartin@physics.ucsb.edu}
\altaffiltext{6}{Astronomy Department 
\& Theoretical Astrophysics Center, 601 Campbell Hall, 
The University of California, Berkeley, CA 94720; 
eliot@astro.berkeley.edu}
\altaffiltext{7}{Princeton University Observatory, Peyton Hall,
  Princeton NJ 08544-1001; thomp@astro.princeton.edu}
\altaffiltext{8}{Lyman Spitzer Jr.~Fellow}

\begin{abstract}
Roughly $80\%$ of Ultraluminous Infrared Galaxies (ULIRGs) show blue
shifted absorption in the resonance lines of neutral sodium,
indicating that cool winds are common in such objects, as shown by
Rupke et al and by Martin. The neutral sodium (NaI) columns indicated
by these absorption lines are $\sim10^{13}-3\times10^{14}\cm^{-2}$,
while the bolometric luminosity varies by a factor of only four. We
show that the gas in ULIRG outflows is likely to be in photoionization
equilibrium. The very small ULIRG sample of Goldader et
al. demonstrates that the ratio of ultraviolet flux to far infrared
flux varies by a factor $\sim100$ from object to object. While the
Goldader sample does not overlap with those of Rupke et al. and
Martin, we show that such a large variation in ultraviolet flux will
produce a similar variation in the column of neutral sodium for a
fixed mass flux and density. However, if the cold gas is in pressure
equilibrium with a hot outflow with a mass loss rate similar to the
star formation rate, the range of ionization state is significantly
smaller. Measurements of the UV flux for objects in the Martin and
Rupke et al. catalogs will definitively determine if photoionization
effects are responsible for the wide variation seen in the sodium
columns. If they are, a determination of the gas density and mass loss
rate in the cool winds will follow, with attendant improvements in our
understanding of wind driving mechanisms and of the effects of
galaxies on their surroundings.

\end{abstract}


\section{INTRODUCTION}
Recent observations of the sodium resonance line (NaD) doublet in
Ultra-luminous Infra-red Galaxies (ULIRGS) by Rupke and coworkers
\cite{2002ApJ...570..588R,2005ApJ...631L..37R,2005ApJS..160...87R,2005ApJS..160..115R}
and by Martin (2005; 2006) show that cool outflows from ULIRGS are
common. Martin (2005) finds that 15 of 18 ULIRGS ($83\%$) possess such flows,
while Rupke et al. (2005b) find a detection rate of $80\pm7\%$ for 30
local ULIRGs. These results indicate that the outflow emerges in most
directions. The absorption troughs, which typically extend over
$\sim300\kms$, are not black, which at first blush might suggest that
the outflows are optically thin in the NaD line. However, the doublet
line ratio, which is equal to 2:1 in optically thin gas, rarely has
that ratio in Martin's sample, while Rupke et al. (2005a) find optical
depths for the weaker of the doublet lines that range from $0.06$ to
$7$, with an average of $1.5$. This finding shows that some sight
lines to ULIRG galaxies are optically thick in the NaD line, that the
optically thick outflow covers only about $25-30\%$ of the (optically
emitting) galaxy, but that this optically thick component is seen
toward $\sim80\%$ of ULIRGS. Given the patchy nature of the
interstellar medium in most galaxies, the last finding is not entirely
surprising.

\begin{figure}
\begin{center}
\plotone{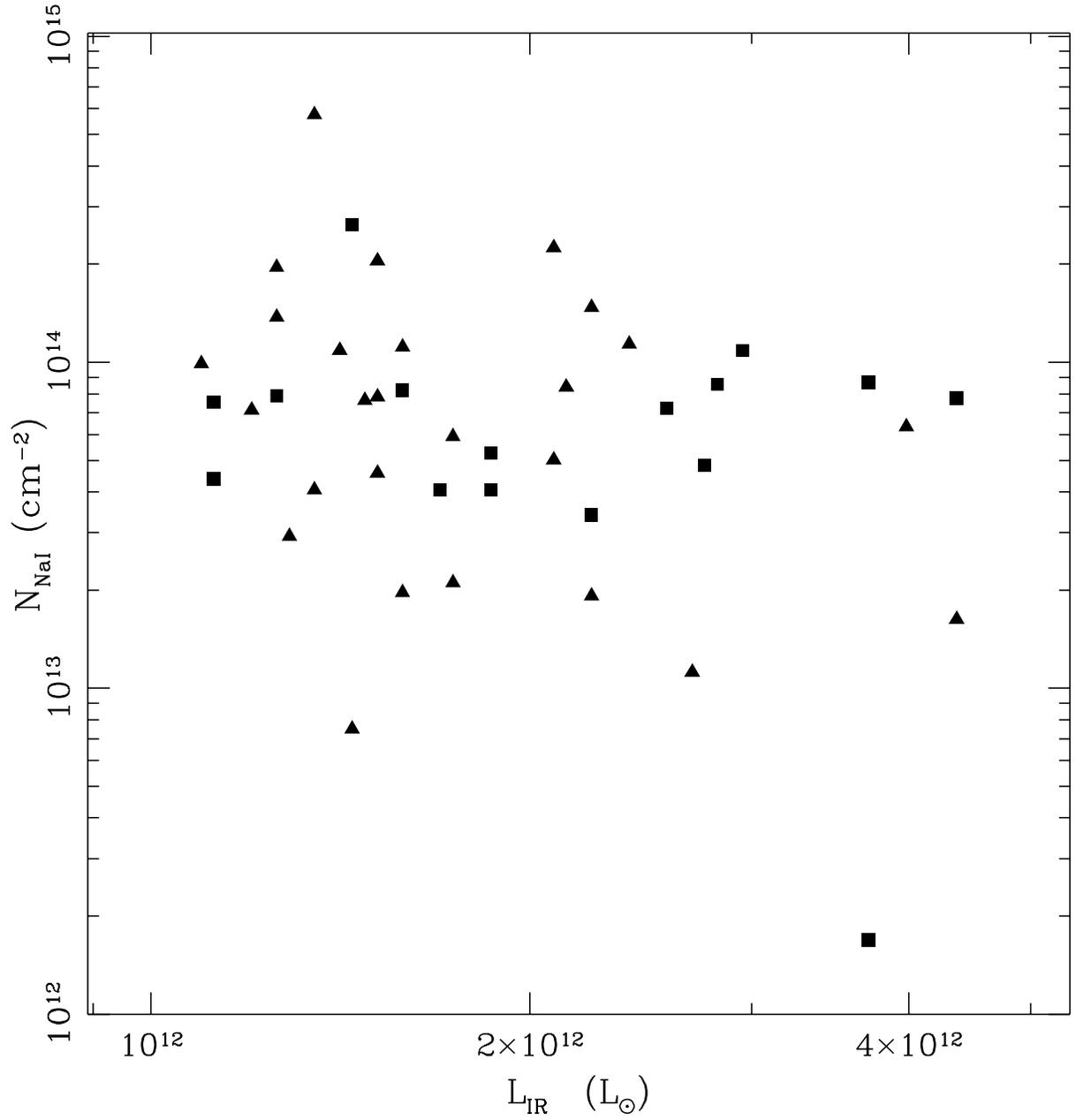}
\end{center}
\caption{Neutral sodium ($NaI$) column as a function of galactic infrared
  luminosity. Data taken from Rupke et al. (2005a) (triangles) and Martin
  (2006) (squares). The Martin results are for sightlines toward the
  respective galactic centers. Note the large spread in the column for galaxies of
  approximately the same infrared luminosity.}
\label{Fig:column}
\end{figure}

However, another implication of the observations is very
surprising. As Figure \ref{Fig:column} shows, the mean column averaged
over both samples (Rupke et al. 2005a; Martin 2006) is $13.7$ in the
log, with a standard deviation of $0.66$ dex, or from
$10^{13}\cm^{-2}$ to $3\times10^{14}\cm^{-2}$. The actual Na I columns
vary from $7\times10^{12}\cm^{-2}$ to $5\times10^{14}\cm^{-2}$ in the
sample of \cite{2005ApJS..160...87R}. Since the luminosity, and star
formation rate of the galaxies in both samples varies by only a factor
of about 4 (the standard deviation is a factor of 2), one might expect
the $N_H$ column and hence the $N_{NaI}$ column toward the galaxies to
vary by a similar factor; in simple galactic wind models the mass
outflow rate increases with increasing star formation rate.

The Na I column can be used to find the mass loss rate from each
galaxy, in principle. For example, in a smooth flow, the density is
related to the mass loss rate by
\be \label{eq:mdot}
\dot M_w = \Omega r^2\mu(r) n(r) v(r),
\ee 
where $\mu(r)$ is the mean molecular weight, $n(r)$ is the number density of
hydrogen, $v(r)$ is the velocity of the flow at a distance $r$
from the center of the galaxy, and $\Omega\le4\pi$ is the global wind cover
factor in steradians. For simplicity we employ a spherical
model, although ULIRGs appear to form a substantial fraction of their
stars in a kiloparsec (or smaller) scale disk \cite{1991ApJ...378...65C}.

The hydrogen column through such a wind is
\be 
N_H = {\dot M_w\over \Omega}\int_{r_0}^\infty {dr\over r^2 \mu(r) v(r)}.
\ee 
A very rough estimate is found by taking $\mu(r)=m_p$ and
$v(r)=v_\infty$, where the latter is the terminal velocity of the
outflow, which in the spirit of approximation adopted here is taken to
be the maximum observed blue shift:
\be \label{eq:column}
N_H\approx{\dot M_w\over\Omega m_p v_\infty r_0},
\ee 
where $r_0$ is the radius at which the absorption line forms. Solving for
the mass loss rate in terms (as far as possible) of observed quantities,
\be 
\dot M_w \approx \Omega m_p r_0 v_\infty N_{NaI} 
\left({N_{Na}\over   N_{NaI}}\right)
\left({N_H\over N_{Na}}\right)d_{Na},
\ee 
where $N_{NaI}$ is the (observed) column of gas phase neutral sodium,
$N_{Na}$ the total gas phase sodium column, $N_H$ is the total
hydrogen column, and $d_{Na}$ is the ratio of total to gas phase Na;
it is larger than unity since much of the Na is locked up in dust
grains. Neither of the two terms in parentheses nor $d_{Na}$ are
measured, nor is $r_0$ well constrained. For solar abundances
$N_H/N_{Na}=4.9\times10^5$ \cite{AG}, while
\cite{1996ARA&A..34..279S} find that the gas phase abundance of Na in
the cool gas of the diffuse cloud toward $\zeta$ Ophiuci is a factor
of $d_{Na}=8.9$ smaller than the solar abundance of Na.  There is
evidence of dust in ULIRG outflows; first, NaD is a resonance line, so
that it scatters continuum photons rather than destroying them. Since
the winds are nearly spherical (they are seen toward 80\% of ULIRGS),
any photons removed from our line of sight by the NaD transition
should appear along some other line of sight as redshifted emission;
since we see no such emission, some other mechanism is removing the
scattered photons, absorption by dust being the number one
suspect. Second, there is a correlation seen between reddening and NaD
absorption toward luminous infrared galaxies, again suggesting dust in
the outflow \cite{1995ApJS...98..171V}; \cite{2000ApJS..129..493H}. In the
rest of this paper we will adopt solar metalicities and the Milky Way
ISM $d_{Na}$ just quoted. Using these values, the typical observed
column $NaI=10^{14}\cm^{-2}$ would correspond to a hydrogen column
$N_H\approx10^{21}\cm^{-2}$ if somewhat less than half the gas-phase
sodium were neutral.

Any local heating, such as might be produced by shocks, will lead to
lower depletion levels, but by less than the factor of ten
corresponding to all the grains being destroyed, given the evidence
for dust cited above.

The force responsible for expelling the wind is uncertain. One
candidate is ram pressure from outflowing supernova-heated gas, which
will entrain cold gas from the interstellar medium of the ULIRG. A
variant on this is ram pressure from hot gas produced by an accreting
massive black hole (an active galactic nucleus, or AGN). The Compton
temperature of both Seyfert and Quasar nuclei is around $10^7$K,
similar to that from supernova heating. Momentum driving from
supernovae can be significant, if most supernovae explode while
surrounded by cold gas. Another candidate for momentum driving is radiation
pressure on dust grains embedded in the cold gas. The radiation could
arise either from stars or from a central black hole. The radiation
pressure from a black hole can exceed that from a starburst. Whether
such a centrally driven outflow can produce an outflow that will be
detected from $80\%$ of the sky is far from certain.

In fact it seems likely that both energy and momentum driving
operate. ULIRGS are enveloped in x-ray emitting gas, Arp 220 being an
excellent example \cite{2003ApJ...591..154M,2002ApJ...581..974C}; the
presence of hot gas is a necessary (but not sufficient) condition for
energy driving of a cool outflow.  As we have just noted, the winds
are most likely dusty. The wind optical depth to optical or UV
emission is likely substantial, while the galaxies themselves are
optically thick even to the far infrared. The outflow may well be
multiphase; the questions to be addressed revolve around which phase
carries more mass, momentum, energy, and metals into the surrounding
intergalactic medium.

To answer such questions, we would like to measure the mass loss rate in
cold gas. If the NaI columns are taken at face value, and the ratio
$N_H/N_{NaI}$ is assumed to be constant, the mass loss rates in cold
gas vary from ULIRG to ULIRG by a factor of $\sim50$. Is this really
true?

There are numerous possible reasons for the NaI column seen in
galactic outflows to vary from object to object. The mass loss rate
will vary from object to object, so that the total column of hydrogen,
and hence Na, will vary, even at a fixed metallicity. The metallicity
of the outflowing gas may vary from object to object. Both these
properties are likely to depend on the star formation rate (and
history); the star formation rate, at least, varies over only a small
range for the samples we are discussing. A more likely cause of
variation in NaI column is a variation in either or both the depletion
on to dust grains and the ionization state of Na. Neutral sodium is a
rather delicate atom, easily ionized by ultraviolet radiation. The
wind may be irradiated either by starlight from the host galaxy, or by
shocks in the wind.

In this paper we argue that the outflow is in photoionization
equilibrium, except possibly for the most UV dim systems, and then
show that if Martin and Rupke et al.'s galaxies have spectral energy
distributions like those of other nearby ULIRGs that have been
observed in the ultraviolet, then it may be that the mass loss
rates are more nearly equal than the observed NaD columns imply;
unless the neutral gas is very dense, $n\sim 10^5\cm^{-3}$, the
ionization fraction, or the ratio of neutral to total sodium,
$(NaI/Na)$, will vary by a large factor from object to
object. This follows from the observed variation of $\sim100$ in the
ratio of UV to FIR flux seen in nearby ULIRGs (Goldader et al. 2002),
and simple photoionization calculations, as we show below.

\section{NEUTRAL SODIUM GAS ABUNDANCES}

The ionization state of the wind depends on the gas density, the 
ionizing flux, and the gas temperature. We start by estimating the gas
density.  We will use canonical values for the luminosity
$L_{FIR}=10^{12}L_\odot$ and terminal wind velocity
$v_\infty=300\kms$. 

To obtain a rough lower limit for the density we will assume a
smooth dust driven wind, which captures the entire momentum output of
the galaxy along a particular sight line. Following Rupke et
al. (2005b) we let $C_\Omega$ denote the fraction of sight lines that
pass through a wind and $C_f$ denote the local cover factor of NaI gas
(along a sight line through the wind), so that $\Omega/4\pi=C_\Omega
C_f$. Rupke et al. (2005b) find $C_\Omega=0.8$ as noted above, and
$C_f\approx0.4$ (their table 2) for local ULIRGs, values which we will
adopt. In that case the
mass loss rate is given by
\be \label{eq:FIR}
\dot M_w={L_{FIR}\over c v_\infty}C_\Omega C_f.
\ee 
For our canonical numbers, this yields a mass loss rate of $\dot
M_w\approx22M_\odot\yr^{-1}$. The star
formation rate needed to power the bolometric luminosity
($10^{12}L_\odot$) is $\dot M_*=L/(\epsilon
c^2)\approx90M_\odot\yr^{-1}$, where $\epsilon\approx8\times 10^{-4}$
for a Salpter IMF with stellar masses between $0.1$ and
$100M_\odot$. The ratio $\eta$ of mass outflow rate to star formation rate is
\begin{eqnarray} \label{eq:eta} 
{\dot M_w\over \dot M_*}&=&C_\Omega C_f \epsilon {c\over
  v_\infty}\nonumber \\
&\approx&0.256
\left({C_\Omega\over 0.8}\right)
\left({C_f\over 0.4}\right)
\left({\epsilon\over 8\times10^{-4}}\right)
\left({300\kms\over v_\infty}\right).
\end{eqnarray} 
Rupke et al. (2005b), using a sample of 30 low redshift ULIRGS, find
$\eta=0.19^{+0.5}_{-0.1}$; the error bars do not include the large
uncertainty in the NaI/Na ratio.

Estimates of mass loss rates in supernova driven winds also tend
to yield results comparable to the star formation rate, e.g.,
\cite{Martin99}, although it is unclear if the hot gas actually
escapes from the galaxy in massive galaxies.

Combining equations (\ref{eq:mdot}) and (\ref{eq:FIR}) yields the estimated density
\be \label{eq:density analytic}
n(r,v;L_{FIR})={L_{FIR}\over 4\pi \mu v_\infty cr^2 v(r)}.
\ee 
Using the canonical values quoted above, $\mu=1.38m_p$, and $r=500\pc$,
we find
\be \label{eq:density} 
n\approx6\left({L\over 10^{12}L_\odot}\right)\left({300{\rm km/s}\over
  v_\infty}\right)
\left({100{\rm km/s}\over v}\right)
\left({500\pc\over r}\right)^2\cm^{-3},
\ee 
We have chosen $r=500\pc$ since this is (roughly) the size of both CO
($\sim 500\pc$, Solomon et al. 1997) and radio disks ($\sim100\pc$,
Condon et al 1991) in typical ULIRGs. The distance between the disk
and the outermost point of the outflow will depend on age, and is
likely to be much larger than $500\pc$, but most of the column will be
accumulated at small radii, see equation (\ref{eq:column}).  While we
have derived this result for a smooth cool outflow, a hot supernova
driven outflow with the same mass loss rate will have the same
density, as can be seen from equation (\ref{eq:mdot}).

If the gas is clumpy, the relevant density will be higher than this
estimate, for the same mass loss rate. In particular, if the cool gas
is driven out of the galaxy by ram pressure from a hot outflow, the
cool gas should be in or near pressure balance with the hot wind. In
that case,
\be 
n\approx n_h {T_h\over T},
\ee 
where the subscript $h$ denotes supernova heated gas. The hot gas has
$T_h\approx10^7$K, while the cold gas has $T\lesssim100$, as shown by
the numerical results described below. If the mass loss rate in hot
gas is of order the star formation rate, the cold gas has
$n\gtrsim6\times10^5\cm^{-3}$ near the launching radius. This is
slightly higher than the density of the ISM in the inner disks of
ULIRGs. Since the ionization state of the gas depends on the gas
density, it may be possible to distinguish between ram pressure driven
flows and dust driven flows. We will return to this point in \S4.

\subsection{Timescales}
With this estimate of the gas density we are in a position to estimate
several relevant time scales. The first is the dynamical time scale,

\be \label{eq:outflow} 
\tau_{\rm dyn}\approx5\times10^{13}
\left({r\over 500\pc}\right)
\left({300\kms\over v_\infty}\right)
\s.
\ee 

In a clumpy flow, the size of the cold clumps should be much
less than the size of the galaxy. For example, in our own galaxy,
observations of cool gas on top of the hot outflow GSH 242-03+37
(McClure-Griffiths et al. 2005) indicate that the cool gas has a
length scale $l$ of order tens of parsecs, while sitting on hot gas 1,500
parsecs above the galactic plane. The sound crossing time for such
clumps is
\be 
\tau_{h,MW}\approx10^{12}\left({l\over 10\pc}\right)
\left({300\kms\over c_h}\right)s
\ee 
for the surrounding hot gas. These clumps have a column density
$N_H\approx3\times10^{19}\cm^{-2}$, a factor of about ten lower than
the column through the disk of the Milky Way, and a density
$n=N_H/l\sim1\cm^{-3}$ for the observed size $l=10\pc$, a density similar to
that in the disk as a whole.

In order to estimate the time scale for the evolution of the cold
outflowing gas in ULIRGs, we need to estimate the length scale of the
cold gas. To estimate the size of a clump of cold gas driven out of a
ULIRG disk by a hot outflow, we use the inferred $N_H$ for the NaI
carrying outflow, and assume that the initial gas density is equal to
that in the disk midplane, as is the case in the Milky Way;
$l=N_H/n=10^{21}\cm^{-2}/10^5\cm^{-3}=10^{16}\cm$. In Arp 220, the
star forming disk(s) have $r\approx100\pc$ with a disk scale height
$H$ satisfying $H/r\approx0.1$ \cite{1999ApJ...514...68S}, so
$l/H\approx3\times10^{-3}$, a factor of ten or so smaller than ratio
of $0.05$ seen in the Milky Way.

This short length scale has important implications for the pressure
and hence the density in the NaI bearing gas. To work out these
implications, we first show that if the hot gas is responsible for
pushing the cold gas out of the galaxy, the mass loss rate in the hot
gas must exceed that of the cold gas by the factor $4\pi/
\Omega\approx5$. We then show that the density of
the cold gas is $n\gtrsim 6\times10^5\cm^{-3}$ for our
cannonical numbers. Gas of this density exposed to the radiation from
a ULIRG will be neutral, with all the gas-phase Na in the form of NaI.

We assume that the cool gas seen in ULIRG outflows comes in roughly
spherical clumps of radius $l$, with column
$N_H\approx10^{21}\cm^{-2}$. These clumps are pushed out of the galaxy
by ram pressure from the hot outflow, which exerts a force \cite{mqt}
\be  
F_{ram}=m_pn_hv_h^2\pi l^2,
\ee  
where $v_h$ is the velocity of the hot
outflow. This force must exceed the force of gravity,
\be  
F_{grav}={v_c^2\over r}m_pn,
\ee  
where $v_c$ is the circular velocity of the galactic disk; we note
that $v_c\lesssim v_h$. Introducing the cover factor $\Omega_h$ of the
hot gas (which should be close to $4\pi$ since the sound speed of the
hot gas is comparable to the bulk velocity of the gas), we find that
the mass loss rate of the hot gas must exceed
\be  
\dot M_h\gtrsim \left({\Omega_h\over\Omega}\right)
\left({v_c\over v}\right)
\left({v_c\over v_h}\right)
\dot M_w
\ee  
if the hot gas is to accelerate the cold clouds out of the galactic
potential. 

The mass loss rate in hot gas must exeed that in cold gas by the ratio
of the cover factors; we expect that this is roughly the inverse of
$C_fC_\Omega$, or about a factor of five. For ram pressure driving to
be important, the density of the hot gas should be that given by
eqn. (\ref{eq:density}).

The hot gas will
travel accross any embedded cold clump in a time
\be  
\tau_h\approx3\times10^8\left({l\over 10^{16}\cm}\right)
\left({300\kms\over c_h}\right)s,
\ee  
much shorter than the dynamical time.

In a supernova driven flow shocks may be driven into the cold
clouds. Let us suppose that the mass loss rate in hot gas is equal to
the star formation rate (and thus a factor $\sim5$ times the observed
mass loss rate of the cold gas in ULIRGs). The pressure in the cold
clouds is
\be  
P_c=1.38\times10^{-9}
\left({n\over10^5\cm^{-3}}\right)
\left({T\over100\,{\rm K}}\right){\rm dynes}\cm^{-2},
\ee  
while the pressure in the hot gas is
\be  \label{eq:P_hot}
P_h\approx8\times10^{-9}
\left({n_h\over6\cm^{-3}}\right)
\left({T\over10^7\,{\rm K}}\right){\rm dynes}\cm^{-2},
\ee  

This over-pressured (from the point of view of the cold gas) hot gas
moves at its sound speed, which is a factor of 100 or more higher than
the thermal sound speed in the cold gas. A cold gas cloud that is
exposed to the hot gas will experience a six-fold increase in external
pressure in a time of order one percent of the cloud dynamical time,
i.e., it sees what is effectively an instantaneous jump in its
external pressure. The result will be a shock of Mach number
$\sim\sqrt{6}$ propagating into the cloud. If the initial density of
the cold gas is smaller, the shock will be stronger.

These shocks will cross the cloud in a time
\be  
\tau_{s}\approx5\times10^{10}
\left({l\over10^{16}\cm}\right)
\left({v_s\over2\kms}\right),
\ee  
where $v_s=\sqrt{P_h/ m_pn}$ is the shock velocity. Since this is
much less than the outflow time given by equation (\ref{eq:outflow}),
the cold clumps should be in pressure equilibrium with the hot gas, if
the hot gas is dynamically important; this statement is true even for
$l$ a factor of 10 or 100 larger than our fiducial value of
$10^{16}\cm$.  Indeed, in our galaxy, McClure-Griffiths et al. (2005)
conclude that the clumps they see are in pressure equilibrium with the
surrounding hot gas.

We have assumed that the shock is isothermal. In fact,
the cooling time for the postshock gas is:
\begin{eqnarray} 
\tau_{cool}\approx 1.5\times 10^{7}
\left({T_s\over 10^3\K}\right)&&
\left({3\times10^{-26}\erg\cm^3\s^{-1}\over\Lambda}\right)\nonumber\\
&&\left({3\times10^5\cm^{-3}\over n_s}\right)\s,
\end{eqnarray} 
where we have scaled to a post-shock temperature $T_s$ appropriate for
a $2\kms$ shock. The numerical value for $\Lambda$ is from
\cite{1972ARA&A..10..375D}. The cooling length
$\tau_{cool}v_s\approx4\times10^{12}\cm$, much shorter than any other
length in the problem, so the shock is effectively isothermal. The
cooling column, $\tau_{cool}v_sn\approx 10^{18}\cm^{-3}$,
consistent with the general result given by
\cite{1980ARA&A..18..219M}. 
In fact this result is more general. Any shock propagating through
dense gas will cool on a time scale short compared to the outflow time
at $r=500\pc$. The maximum cooling column is \cite{1980ARA&A..18..219M}
\be 
N_{cool}\approx10^{19}-10^{20}\cm^{-3},
\ee 
for $v_s<20\kms$, and less for higher shock velocities.

The minimum value for $\Lambda\approx
10^{-26}\erg\cm^3\s^{-1}$ \cite{1972ARA&A..10..375D} for
$T\gtrsim30$K, so as long as $n>10\cm^{-3}$ the gas will cool to
$30K$ in less than the outflow time.

This result shows that, in the event that the initially cool, dense
component of the ISM is shocked and hence heated in the wind, it will
quickly cool again. Since the ratio of temperatures of the hot and
cold gas exceeds $10^5$, while, as we have just shown, the two are in
pressure equilibrium, the density of the cold gas must exceed
$n=6\times10^5\cm^{-3}$, as we stated above.

If the mass-loss rate in hot gas is less than the star formation rate,
so that some mechanism other than ram pressure is responsible for
expelling the cold gas from the galaxy, $P_h$ will be similar to or
smaller than the pressure in the cold clouds; the only shocks present
will be those driven into the clouds by the ram pressure of the hot
gas as it encounters cold gas. The cold gas will not be pressure
confined by the hot gas, so the cold gas will expand at its sound
speed as it flows out. If the initial length scale is $10^{16}\cm$,
and the mass loss rate is that given in equation (\ref{eq:eta}), the
cold gas will exand to a cover factor $C_f$ of order unity by the time
it is $500\pc$ from its launching radius.

The sodium ions in the post-shock gas will also quickly recombine once
the gas cools, as we show next.  The recombination time for sodium in
cool $T\sim100{\rm K}-1000$K gas is
\be 
\tau_{rec}={1\over \alpha_{Na} n_e}\approx10^{12}
\left({0.06\cm^{-3}\over n_e}\right)\s.
\ee 
In this expression, $\alpha_{Na}$ is the total recombination
coefficient of NaII to NaI. The radiative recombination coefficient
$\alpha_{Na}(T=100{\rm K})\approx10^{-11}\cm^3\s^{-1}$
\cite{1996ApJS..103..467V}, which we have assumed dominates the
recombination. We have scaled to the minimum electron density, for gas
of any density, found in our numerical photoionization calculations
using CLOUDY, described below.  This minimum electron density occurs
in gas with low density ($n=6\cm^{-3}$ and low UV/FIR flux ratios,
$\sim10^{-9}$; the electron density for SEDs with higher UV/FIR flux
density ratios is much larger, leading to shorter recombination times.

For warm ionized gas, $T\approx10^4$K, $n_e\approx n$, while
$\alpha(T=10^4)\approx2\times10^{-13}\cm^3\s^{-1}$. Thus, for any
temperature between $100$K and $10^4$K the recombination time is far
shorter than the dynamical time.  We conclude that shocks are not an
efficient mechanism for maintaining sodium atoms in a singly ionized
state

As the gas flows away from the disk plane the density and pressure in
the hot component will decrease as $1/r^2$, so the density in pressure
confined cold clumps will also decrease, lengthening
$\tau_{cool}$. However the cooling time will remain short compared to
the outflow time, which increases as $r$, out to megaparsec distances.

Even for a smooth flow not in pressure equilibrium with a
massive hot outflow, the density $n\approx6$, so the cooling time is
$\tau_{cool}\approx8\times10^{12}\s$, still short compared to the
outflow time. However, if shocks occur at large enough distances, the cooling
time may exceed the outflow time. Nominally this would occur at
distances of order $5\kpc$, but this neglects the compression of the
gas in the shock.

The photoionization time is
\begin{eqnarray} 
\tau_{ion}&=&{4\pi r^2 \langle h\nu \rangle\over a_{NaI} L_{UV}\hfil
}\nonumber \\
&\approx & 10^5\left({r\over500\pc}\right)^2\left({10^{12}L_\odot\over
 L_{FIR}}\right) \left({L_\nu^{FIR}\over L_\nu^{UV}}\right)\s,
\end{eqnarray} 
where $a_{NaI}\approx10^{-19}\cm^2$ \cite{1996ApJ...465..487V} is the
photoionization cross section and $L_{UV}\approx \nu L_\nu^{UV}$ is
the luminosity at the sodium edge. The photoionization time ranges
between $\approx10^{9}$ and $3\times10^{11}\s$ for observed
UV/FIR flux density ratios (see Figure \ref{Fig:UV} for the flux density ratios).

We have seen that both the recombination time and the photoionization
time are shorter than the dynamical time, at least for temperatures
below $10^5$K, while the post-shock cooling time is shorter still. We
conclude that the NaI laden gas is in photoionization equilibrium.

\subsection{Photoionization equilibrium}
Consider a dilute gas of number density $n$ irradiated by the light
from a ULIRG. We saw above that both the photoionization and
recombination time scales are shorter than the dynamical time
scale, so the ionization state of sodium is controlled
by the flux near $5.139$ eV, the ionization potential of Sodium I, and
by the flux near the Lyman edge at $13.6$ eV.

The Lyman edge flux controls whether the gas as a whole is neutral. In
highly ionized gas, the neutral fraction of hydrogen is given by
\cite{osterbrock}
\be 
{n_0\over n}= {\alpha_H\over a_H c} U^{-1}\approx 10^{-6} U^{-1},
\ee 
where $\alpha_H\approx10^{-13}\cm^3\s^{-1}$ is the recombination
coefficient for hydrogen, $a_H\approx 6\times 10^{-18}\cm^2$ is the
photoionization cross section, $c$ is the speed of light, and 
\be 
U\equiv {L_0/\langle h\nu_0\rangle\over 4\pi r^2 n_e c}
\ee 
is known as the ionization parameter. The rate of emission of
ionizing photons, $L_0/\langle h\nu\rangle$ is 
defined by
\be \label{eq:lyman edge}
{a_{\nu_0}L_0\over \langle h\nu_0\rangle }\equiv \int_{\nu_0}^\infty
d\nu {a_\nu L_\nu\over h\nu},
\ee 
where $\nu_0$ is the frequency at the Lyman edge, and $L_\nu$ is the
flux density (luminosity
per Hertz, $\ergs {\rm \, Hz^{-1}}$), at frequency $\nu$ . 

The Lyman edge luminosity $L_0\sim \nu_0L_{\nu_0}$ is not known for
ULRIGs, but it is likely to be much less than the flux at longer
wavelengths, for two reasons. First, most ULIRGs are powered primarily
by stars, which have large intrinsic Lyman edges. ULIRGs powered by
AGN may have larger Lyman luminosities; if so they should show less
absorption in NaI. Second, the dusty gas both in the galaxy proper and
in the outflows will have a substantial (perhaps dominant) fraction of
neutral hydrogen, which will absorb radiation beyond the Lyman
edge. With our canonical values for mass loss rate (and hence
density), and for expected Lyman edge fluxes
($L_{0}\lesssim10^{42}\ergs$, corresponding to smooth
extrapolations from UV flux {\em density} ratios
$L_{\nu}^0/L_{\nu}^{FIR}\approx 10^{-6}$, see Figure \ref{Fig:UV}), the
neutral fraction of hydrogen is at least $10^{-4}$; the column through
the ULIRG wind is of order $10^{21}\cm^{-2}$, so the wind has an
optical depth of order unity at the edge.

Galaxies with no intrinsic Lyman edge and large UV to FIR flux density
ratios 
\be 
L_\nu^{UV}/ L_\nu^{FIR}>10^{-5}
\ee 
will then have ionized
winds, independent of any possible heating due to shocks or supernova
driving. Such winds will not contain an observable column of neutral
sodium.

In contrast, galaxies that have no intrinsic Lyman edge but which have
flux density ratios less than $10^{-5}$ will have winds that are
optically thick at the edge, as will almost any galaxy with an
intrinsic Lyman edge. In such galaxies the wind will consist primarily
of (nearly) neutral hydrogen, assuming that it is not strongly
heated by supernovae or an AGN. Such galaxies may or may not have substantial
columns of neutral sodium, as we now show.

In gas in which hydrogen is nearly neutral, the neutral fraction of
sodium is given roughly by
\be \label{eq:neutral} 
{n_{NaI}\over n_{NaII}}\approx {\alpha_{NaI}\over a_{NaI} c}{n_e\over n} U_{NaI}^{-1},
\ee 
where
\be  
U_{NaI}\equiv {L_{NaI}/\langle h\nu_{ed}\rangle\over 4\pi r^2 n c}
\ee 
is the ratio of the number density of NaI ionizing photons to the
number density of hydrogen. The subscript $ed$ refers to the sodium
photoionization edge, i.e.. $\lambda_{ed}\approx 0.2\mu{\rm m}$, and the
quantity in angular brackets is the average energy of ionizing
photons. The numerator in this definition is, by analogy with equation
(\ref{eq:lyman edge})
\be 
{a_{\nu_{ed},NaI}L_{NaI}\over \langle h\nu_{ed}\rangle}\equiv \int_{\nu_{ed}}^\infty d\nu
{a_{\nu,NaI}L_\nu\over h\nu}.
\ee 

Combining these expressions, we estimate the neutral fraction of Na as
\be \label{eq:ratio}
{n_{\rm NaI}\over n_{\rm NaII}}={\alpha_{NaI}\over a_{\nu_{ed},NaI}}{4\pi r^2 n_e \langle
    h\nu_{ed}\rangle \over L_{FIR}} {\lambda_{ed}\over \lambda_{FIR}}
{L_\nu^{FIR}\over L_\nu^{UV}}e^{\tau_d},
\ee 

The factor $e^{\tau_d}$ accounts for the possible presence of dust.
We remarked above that the lack of NaD emission lines 
together with the correlation of NaD absorption with reddening
suggests that the outflows are dusty, but, as also noted above, the
outflows do not cover the galaxy. The light that we see is thus a
mixture of starlight directly from the galaxy combined with light
filtered through the dusty wind. The sodium ions, on the other hand,
are embedded in the wind, so they see the starlight after it has been
filtered through the dust in the wind interior to their position. To
account for this, the flux $L_\nu^{UV}$ should be reduced from that
observed by a factor $\sim e^{-\tau_d}$, where $\tau_d$ is some
appropriate fraction of the dust optical depth through the wind. The
hydrogen column through the wind is uncertain (since the conversion
from Na to H columns is uncertain), but if the mass loss rate is of
order the star formation rate the hydrogen column is of order
$5\times10^{21}\cm^{-2}$ (see equation \ref{eq:column}). Assuming a
galactic gas to dust ratio, the dust optical depth would be
$\tau_d\sim 2$ in the UV. Numerically,
\be \label{eq:numerical ratio}
{n_{\rm NaI}\over n_{\rm NaII}}\approx 4\times10^{-9}
{L_\nu^{FIR}\over L_\nu^{UV}}\left({n\over 1\cm^{-3}}\right),
\ee 
where we have taken $\tau_d=1$. In terms of $U_{NaI}$,
\be 
{n_{\rm NaI}\over n_{\rm NaII}}\approx 3\times10^{-4}/U_{NaI}.
\ee 
This expression is valid for $U_{NaI}\gtrsim10^{-3}$. For
values of the ionization parameter less than this, the neutral sodium
fraction will saturate (at a value of order $0.5$, see the numerical
results below).

We will see that if the cool outflow is driven by a hot wind with a mass
loss rate comparable to the star formation rate, and the cool
component has a similar mass loss rate, $U_{NaI}$ ranges from
$10^{-7}$ to $10^{-3}$; for a dust driven outflow with the same mass
loss rate the range is $10^{-3}\lesssim U_{NaI}\lesssim10$.

The object to object variation in the UV flux is most naturally
explained by variations in the dust optical depth along our line of
sight; thus it may be appropriate to set $\tau_d$ to some smaller
value in objects with (relatively) large UV fluxes. On the other hand
it is possible that even in objects with large UV fluxes the hydrogen
column through the wind is still of order $10^{21}\cm^{-2}$, but that
some of the star formation is unobscured; even in the object with the
highest UV flux in the Goldader et al. (2002) sample, the UV flux is
less than a tenth of the bolometric flux. The mass loss rate may or
may not be affected by this bolometrically small UV flux. If it is
not, then $\tau_d\approx1$ would be appropriate even for those objects
with large UV fluxes.

We were not able to find photometry in the UV bands for the IRAS
sources in Martin (2005) or in \cite{2005ApJS..160...87R}. As a proxy,
we used the SEDs for for the luminous infrared galaxies observed by
\cite{2002ApJ...568..651G}. The data, obtained from NED, are shown in
Figure \ref{Fig:UV}. The flux density near the Na I edge (about 0.2
microns) is smaller, relative to the bolometric luminosity, in Arp 220
than in IRAS 15250+3609. The difference is roughly a factor of 90; The
relative Na I edge flux density in VV 114 is another factor of 4 larger than
that in IRAS 15250+3609. The other objects in Goldader et al. (2002),
IRAS 08572+3915, Mrk 273, IC 883, and IRAS 19245-7245, have flux
ratios intermediate between IRAS 15250+3609 and Arp 220. We note that
Mrk 273 is an AGN, with a luminosity between a Seyfert and a
quasar. Despite this fact, it has a modest UV flux.

Since the Na edge flux densities of our  proxy galaxies differ by
a factor of $\sim 90$ (for the same FIR flux) we expect a similar
difference in the fraction of neutral Na in their outflows. For VV
114, with $L_\nu^{UV}/L_\nu^{FIR}\approx10^{-4}$, equation
(\ref{eq:ratio}) predicts a neutral fraction of $\sim
2\times10^{-5}(n/1\cm^{-3})e^{\tau_d}$; for Arp 220, with
$L_\nu^{UV}/L_\nu^{FIR}\approx2\times10^{-7}$, the predicted neutral
fraction is $\sim 0.01(n/1\cm^{-3})e^{\tau_d}$.

\section{NUMERICAL CALCULATIONS}
We have carried out numerical calculations of the photoionization
state of gas irradiated by sample ULIRG spectral energy distributions
(SED). These calculations were performed with version 06.02.09a of
Cloudy, last described by Ferland et al. (1998). For the sake of
simplicity we modeled the wind using a constant density slab, with
six different densities ($6$, $60$, $600$, $6000$, $6\times10^4$ and
$6\times10^{5}\cm^{-3}$), with the lowest and highest densities
corresponding to a smooth cold outflow and to a clumpy cool medium in
pressure equilibrium in a hot outflow with mass loss rate equal to the
star formation rate. The intermediate densities model clumpy cold outflows
not driven by hot gas. Dust grains were
included in the model; we used the ISM option of cloudy, which
depletes the gas abundances so that the total metallicity, gas plus
dust, is solar. The dust and gas are mixed in the slab, which lies
between the continuum source and the observer.

For each density the cold gas is illuminated by various SEDs, all having
a total luminosity of $10^{12}L_\odot$, appropriate for ULIRGs. Three
of the SEDs employed were those of VV 114, IRAS 15250+3609, and Arp
220. In addition we calculated a series of models based on the SED of
Arp 220. In these models we adjusted the flux density at 5.1eV
($\approx 2400$ \AA), varying the FIR to UV ratio from
$L_\nu^{UV}/L_\nu^{FIR}=10^{-9}$ to $10^{-4}$. In doing so we
fit a power law from the last IR data point to $2400$ \AA. From $2400$
to $1450$ \AA~we assumed that $\nu L_\nu$ was flat, i.e., $L_\nu\sim
\nu^{-1}$ as in the 2kpc data of Goldader et al. (2002) for Arp
220. From $1450$ \AA~ to the Lyman edge we simply continued the
spectrum with the same slope, $L_\nu\sim\nu^{-1}$.

We note that the UV slope for many of the galaxies in Goldader et
al. (2002) varies with the aperture size, and that the total flux
scales roughly with the square of the aperture size, the latter
indicating that the UV flux emerges from a much larger region than
the radio or CO emission. If the same holds true for the ULIRGs in
general, the UV flux will not decrease as rapidly with distance from
the galactic center as will the density in a smooth outflow. The
result will be an increasing ionization parameter with increasing
radius; if the cool gas is clumpy, and not pressure confined, then the
density will not decrease as $1/r^2$, and the ionization parameter
will not increase as rapidly as in a smooth wind. We have not
attempted to model such effects.

For models with no intrinsic Lyman edge, the spectrum from $13.6$ eV
to the lone x-ray point at $1$ KeV was a simple power law. In models
with an intrinsic Lyman edge, the spectrum from the edge to the x-ray
point was also a power law, but with a discontinuity at the energy
corresponding to the Lyman edge.

We explored other SEDs.  We found that the details of the continuum
shortward of $2400$ \AA~do not matter as long as the Lyman edge flux
is small enough; if the fraction of neutral hydrogen is larger than
about $10^{-4}$ ($U<10^{-2}$) at the face of the slab, the wind is
optically thick at the Lyman edge and the NaI fraction is controlled
by the flux at the NaI edge. If this condition on the neutral hydrogen
fraction is not met, there is essentially no neutral sodium unless an
unphysically large column ($N_H>10^{22}\cm^{-2}$), sufficient to
produce a Lyman edge, is used.

Calculations with CLOUDY verify the simple arguments made in the
previous section; the neutral fraction of Na can vary by factors of
100 or more due to variations in the SED of the host galaxy. The
neutral fraction varies from a maximum near $1$ for an SED with a UV
to FIR flux density ratio of $10^{-8}$ to a low of $10^{-4}$ for the
spectrum of VV 114 (assuming a substantial Lyman break; if there is no
break, the neutral fraction of Na can be much smaller).

\begin{figure}
\begin{center}
\plotone{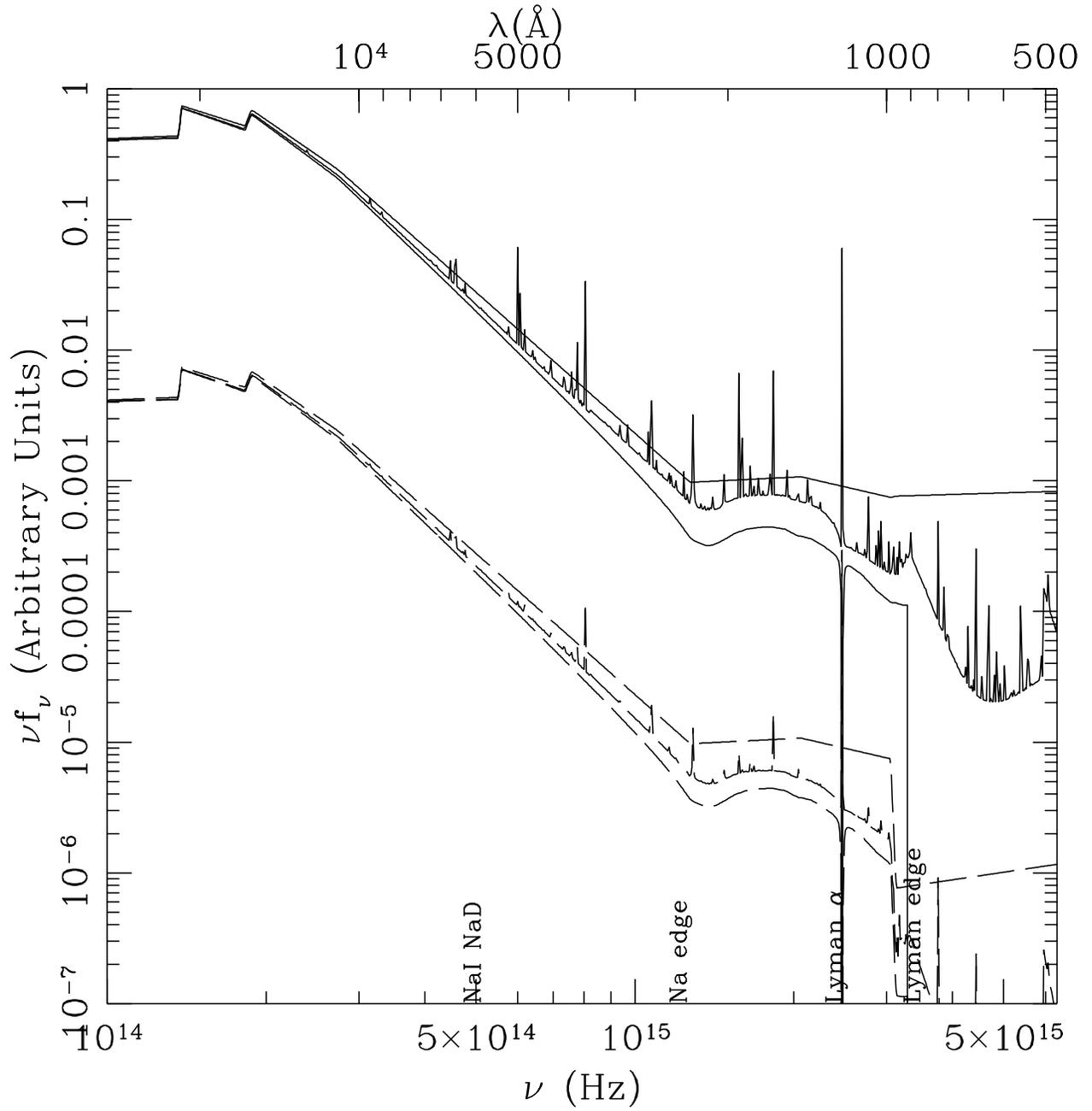}
\end{center}
\caption{ Input spectra of a CLOUDY model with a UV to FIR flux
  density ratio of $10^{-7}$ and no Lyman break (top solid lines) and
  with a Lyman break (top dashed curve); the latter has been displaced
  downward by a factor of one hundred to clarify the presentation. In both
  cases the middle curve shows the transmitted plus emitted spectrum,
  while the bottom curve shows the transmitted spectrum alone. }
\label{Fig:cloudy sed}
\end{figure}

Figure \ref{Fig:cloudy sed} shows the input spectrum of a CLOUDY
run, the spectrum transmitted through the outflow, and the sum of the
transmitted and emitted spectrum. The UV to FIR flux density ratio is
$10^{-7}$ at the face of the slab, and there is no Lyman edge. After
passing through the slab, the spectrum exhibits a strong Lyman edge,
and strong resonance lines, including Lyman alpha. The effect of dust
attenuation can be also seen clearly; this attenuation enhances the
fraction of Na I over what is seen in calculations with no dust. The
dashed lines show a similar calculation in which an intrinsic Lyman
edge appears in the input spectrum (which has been shifted downward to
distinguish the two cases). Note the lack of prominent emission lines,
since there are few ionizing photons to power such emission.

\begin{figure}
\begin{center}
\includegraphics[height=15cm,width=15cm]{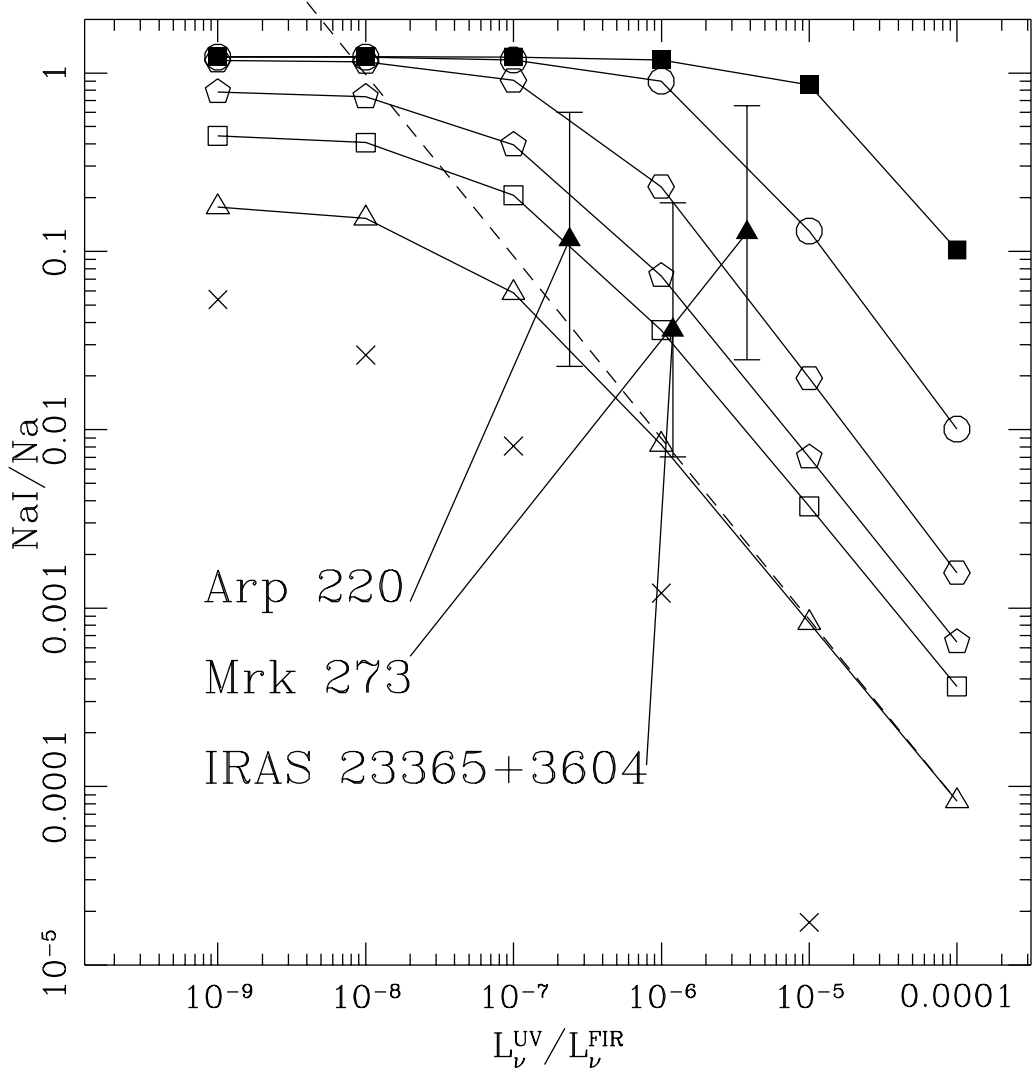}
\end{center}
\caption{ Neutral sodium fraction, $N_{NaI}/N_{Na}$, as a function of
 $L_\nu^{UV}/L_\nu^{FIR}$. The open polygons are calculated using the
 SED of Arp 220, but with the near UV data points altered to give
 various ratios of $L_{UV}/L_{FIR}$.  The (assumed constant) density
 of the absorbing gas is $n=6\,\cm^{-3}$ (open triangles),
 corresponding to a smooth wind with $\dot M_w=22 M_\odot/\yr$ at a
 distance $r_0=500\pc$ from the source, $n=60\,\cm^{-3}$ (open
 squares), $n=600\,\cm^{-3}$ (pentagons), $n=6000\,\cm^{-3}$
 (hexagons), and $n=6\times10^4\cm^{-3}$ (circles). The filled squares
 give the ionization fraction for $n=6\times10^5\,\cm^{-3}$,
 corresponding to cold gas in pressure equilibrium with a hot outflow
 with a mass loss rate equal to the star formation rate. The crosses
 correspond to $n=6\,\cm^{-3}$ and various flux density ratios, using
 SEDs with no Lyman edge. The dashed line is
 the prediction of equations (\ref{eq:column}) and (\ref{eq:numerical
 ratio}) for $n=6\cm^{-3}$ for an SED with a Lyman edge. The solid
 points correspond to the measured $NaI$ column for Arp 220, Mrk 274,
 and IRAS 23365+3604, under the assumption that $\dot M_w=\eta\dot
 M_\odot$, with $\eta=0.25$, and $r_0=500\pc$.  }
\label{Fig:neutral_column}
\end{figure}

Figure \ref{Fig:neutral_column} shows the ratio $N_{NaI}/N_{Na}$ as
a function of the UV (evaluated at $2000$ \AA) to FIR (100 micron)
flux density ratio for a range of densities. We have assumed
that the flux shortward of the Lyman edge is a factor of 10 lower than
on the long wavelength side. From the last UV data point to the edge
we assumed a power law of the form $L_\nu\sim \nu^{-1}$; from just
shortward of the edge to the first x-ray data point we assumed a
simple power law.

The open points correspond to the artificial SEDs based on the SED of
Arp 220, assuming a Lyman edge. The density is held constant at the
value $n=6\,\cm^{-3}$ (open triangles), $n=60\,\cm^{-3}$ (squares),
$600\,\cm^{-3}$ (pentagons), $6000\,\cm^{-3}$ (hexagons),
$6\times10^4\,\cm^{-3}$ (circles), and $6\times10^5\,\cm^{-3}$
(filled squares); the open triangles corresponds to a completely
smooth wind with $\dot M_w=22M_\odot/{\rm yr}$, the filled squares to
pressure confined cold clouds in hot gas with a mass loss rate equal
to the star formation rate. 

The crosses correspond to the artificial Arp 220 SEDs, assuming no
Lyman edge, and a smooth wind. The lower neutral sodium column in the
no-edge case compared to the intrinsic-edge case arises as follows:
Lyman continuum photons penetrating the inner face of the slab keep
hydrogen ionized until the optical depth at the Lyman edge reaches
unity; beyond that depth hydrogen recombines, and the neutral fraction
of sodium is essentially the same as in the case of an SED with an
intrinsic Lyman edge. When the ``Stromgren depth'' is larger than the
column in the wind, as when the UV/FIR ratio is $10^{-4}$, the neutral
fraction of sodium is tiny ($\sim10^{-9}$), so the corresponding cross
in Figure \ref{Fig:neutral_column} is off the bottom of the plot.

The dashed line in Figure \ref{Fig:neutral_column} gives the
prediction of equation (\ref{eq:ratio}), which agrees well
with the more complete calculation for UV/FIR ratios larger than
$\sim10^{-7}$, in the case of an intrinsic Lyman edge.

The three filled triangles correspond to the measured
values of NaI for Arp 220, Mrk 273 (data for both taken from Heckman
et al. 2000), and IRAS 23365+3604 (NaI column from Martin 2005), as
labeled in the figure. In each case we assumed that the mass loss rate
from the host galaxy was given by equation (\ref{eq:eta}), then use
the observed NaI column to find the ionization fraction. The error
bars reflect only the variation in $\eta$ found by Rupke et al
(2005b), i.e., only the apparent variation in the mass loss rate for
objects of the same luminosity. To estimate the (unknown) flux at the
NaI edge for IRAS 23365+3604 we extrapolated from the nuclear (2.5
kpc) U-band measurement of Surace \& Sanders (2002), assuming a power
law extending from the last two optical data points. This
extrapolation is highly uncertain, as a glance at Figure \ref{Fig:UV} shows.

With only three data points, those corresponding to Arp 220, Mrk 273,
and IRAS 23365, the interpretation of
Fig. \ref{Fig:neutral_column} is uncertain. In particular, there is no
clear trend of Na column with UV flux. The three data points
correspond to densities in the range $60-6000\cm^{-3}$, suggesting that
the flow is slightly clumpy, but not in pressure equilibrium with a hot flow of
the same total mass loss rate. Clearly more measurements of near UV
fluxes for galaxies in either or both the Rupke et al. and Martin
samples are called for.

\begin{figure}
\begin{center}
\plotone{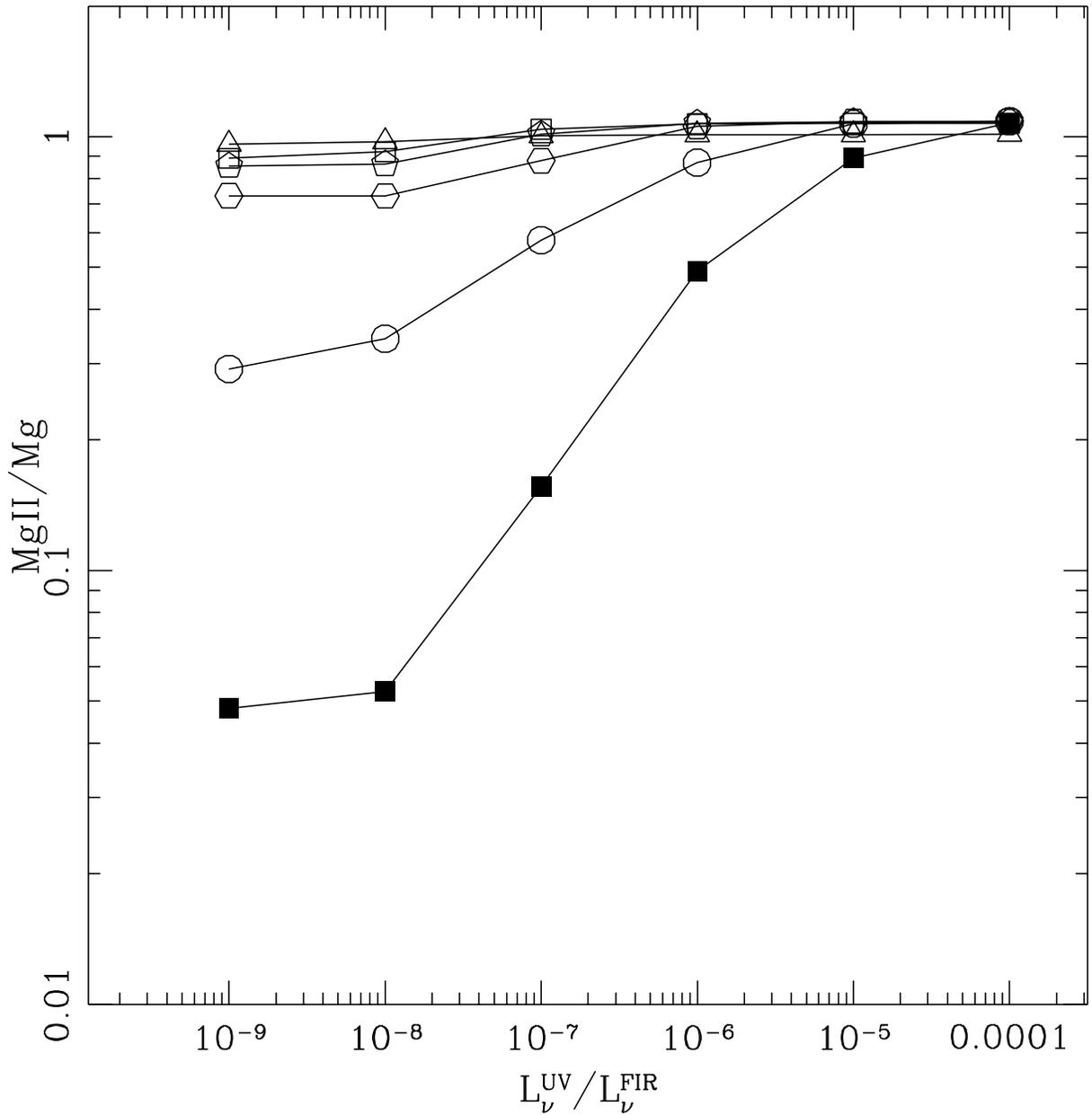}
\end{center}
\caption{ The ratio of Magnesium II to total Magnesium,
  $N_{MgII}/N_{Mg}$, as a function of $L_\nu^{UV}/L_\nu^{FIR}$. The
  SED's and symbols are as in Fig. 4. The bulk of the magnesium is
  singly ionized for all but the highest densities plotted. The column
  of MgII toward a ULIRG would be a {\em decreasing} function of
  $L_{UV}/L_{FIR}$, particularly for pressure confined high density
  clouds. This behavior is the opposite of that predicted for the
  NaI/Na ratio for similarly high density gas.}
\label{Fig:magnesium}
\end{figure}

Figure \ref{Fig:magnesium} is similar to
Figure \ref{Fig:neutral_column}, except that the ion fraction plotted
is that for Magnesium II over the total Magnesium abundance. The bulk
of the magnesium is singly ionized for $n<6\times10^4\cm^{-2}$; even
for higher densities most of the atoms are singly ionized for UV to
FIR flux density ratios larger than $10^{-5}$. The near-UV resonance
transition of MgII should be a good tracer of neutral outflows; if the
MgII column is seen to anti-correlate with UV flux, while the NaI
column does not vary with UV, it would suggest that the cool gas has a
high density and is therefore probably driven out by hot gas. If the
MgII column does not vary with UV flux, while the NaI column does,
then the gas is most likely low density, and not driven out of the
galaxy by hot gas.

\section{DISCUSSION}
We have identified two plausible sources of variation in the
ionization fraction of NaI toward ULIRGs, variations in the (NaI)
ionizing radiation, and global object-to-object variations in gas
density. The limited observations of UV fluxes from ULIRGs
(essentially the objects in Goldader et al. 2002) suggest that the
former source of variation may well be happening.

Variations in the globally averaged density of outflowing gas
translate into variations in filling factor of dense gas if we assume
that $\dot M_w$ does not vary much from object to object. The filling
factor might vary, if, for example, the cold gas were in pressure
equilibrium with hot gas, but the mass loss rate of hot gas varied
from ULIRG to ULIRG. Another possible source of variation in filling
factor might be global differences in the outflow, leading to stronger
or weaker shocks and hence higher or lower post-shock
densities. However, because the sound crossing times of the clumps are
smaller than the outflow time, this source of variation does not
appear to us to be very promising---the clumps would simply re-expand
until they reach pressure equilibrium.

Another possible source of variation is variable depletion onto dust
grains. If high velocity shocks pass through the gas they will destroy
the dust grains, so that no Na is depleted onto dust. As we have
indicated above, there is observational evidence that dust survives in
the outflows, so perhaps any shocks that have passed through the gas
are not strong. On the other hand, if the hot gas has a high pressure,
the density of the cold gas may well be high enough to regrow grains,
or to cause further depletion of Na onto any grains that survive. We
have not attempted to work out the physics of these processes.

The referee kindly pointed out that observations of NGC 6240
\cite{2003AJ....126.2185V} show emission lines consistent with shock
ionization. As we noted above, the column of the shock-illuminated and
-ionized gas is small compared to the total column in the outflow. The
emission lines may be dominated by shock ionization, but it does not
follow that the bulk of the gas in the wind is. A clear way to tell
the difference between photoionized and shock-ionized flows is to look
for a systematic variation of NaI columns with UV flux. If such a
correlation is found, the gas is photoionized. If there is no such
correlation, then it is shock ionized. (The  scaling of UV
flux with aperture size found by Goldader et al. 2002 shows that in
their galaxies at least, the UV emission is not dominated by that
produced in shocks in a wind).

\section{CONCLUSIONS}
The mass loss rates of ultraluminous infrared galaxies, as determined
by observations of the NaD lines, are believed to be substantial---of
order one fifth of the star formation rate.  This estimate assumes
that the ionization fraction of sodium is $\sim0.1$.  The gross
physical properties of ULIRGs, including their stellar masses, gas
masses, sizes, and star formation rates vary by factors of only a few,
so at first glance the assumption of fixed NaI ionization fraction
appears reasonable. However, one property of ULIRGS that appears,
based only on a few galaxies, to vary by a large amount is the
ultraviolet flux. We have shown that cool ULIRG winds are likely to be
in photoionization equilibrium. We used a simple argument to show that
the large object-to-object UV flux variations seen in the small number
of galaxies that have been observed in the ultraviolet would be enough
to produce similarly large variations in NaI column density, if the
density of the cool gas is $n\lesssim10^5(500\pc/r)^2\cm^{-3}$. More
complete photoionization calculations using CLOUDY produce the same
magnitude of variation in NaI/Na with variation in UV flux. They also
show, for these moderate densities, that the MgII column should not
vary appreciably with changes in UV flux.

If, however, the cold gas is in pressure equilibrium with a hot wind
having a mass loss rate similar to the star formation rate of a
typical ULIRG, the cool gas will have a density $n\gtrsim
10^6(500\pc/r)^2)\cm^{-3}$. In that case the ionization parameter is
$U_{NaI}<10^{-3}$, and there will be little variation in NaI column
from object to object. As a check, the (so far unobserved) MgII columns
are predicted to show variations that are anti-correlated with the
UV/FIR ratio.

We conclude that the observed factor of $\sim100$ variation in the
UV/FIR flux density ratios in ULIRGs with similar bolometric
luminosities is consistent with the factor of $\sim40$ variations in
the column density of Na I outflows seen by Martin (2006) and
\cite{2005ApJS..160...87R}; the different UV fluxes may well be the
cause of the variation seen in the NaI columns, if the cool gas has a
density $n\lesssim 10^5\,\cm^{-3}$. A tight correlation between Na I column
and UV flux would indicate that the cool gas is not accelerated by a
hot wind.

The origin of the object-to-object variation in the UV/FIR ratio is not
clear. The known correlation between reddening and NaI column in lower
luminosity galaxies suggests that dust in the outflow may play a
role. However, the NaI optical depth toward luminous infrared galaxies in
the sample of Rupke et al. (2005b) actually exceeds that toward their
(more luminous) ULIRGs. It is unlikely that the UV/FIR ratio is
smaller in the less luminous objects. Thus dust absorption in the
outflow is probably not the sole reason for the low UV fluxes in ULIRGS.

\begin{figure}
\begin{center}
\plotone{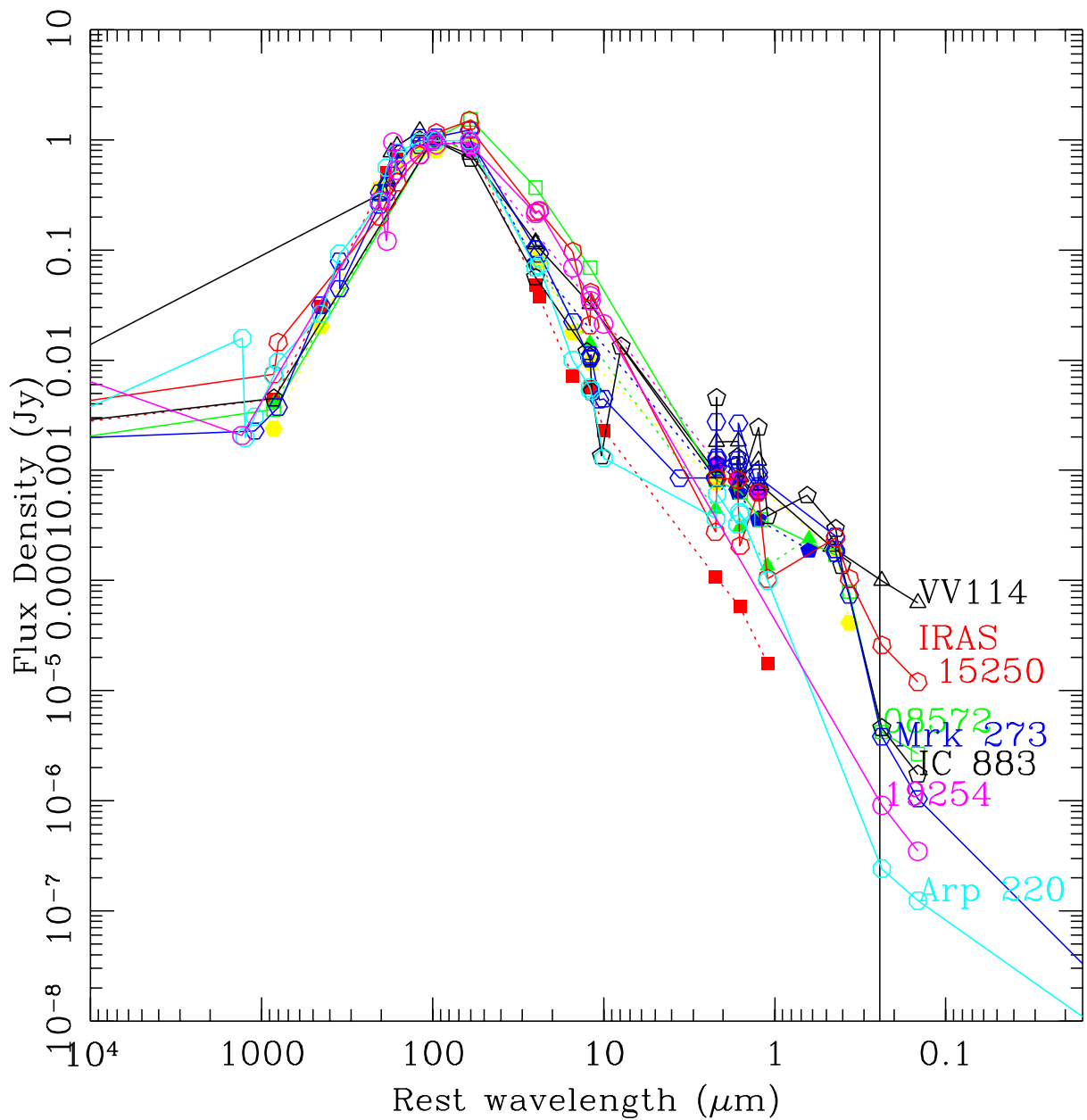}
\end{center}
\caption{Spectral energy distributions of various ULIRGs, normized to
  unity at $100\mu{\rm m}$; data from the NASA/IPAC Extragalactic
  Database (NED), supplimented by Goldader et al. (2002) (we plot the
  $2\kpc$ aperture data) and Martin (2005). The solid symbols
  represent objects from Martin (2005); triangles, IRAS 10565+2448;
  squares, IRAS 17208-0014; pentagons, IRAS 19297-0406; hexagons, IRAS
  23365+3604; septagons, IRAS 00153+5454. The open symbols are from
  objects with HST UV observations by Goldader et al. (2002);
  triangles, VV 114; squares, IRAS 08572+3915; pentagons, IC 883;
  hexagons, Mrk 273; septagons, IRAS 15250+3609; octagons, Arp 220;
  and circles, IRAS 19254-7245. The UV/FIR flux density ratio varies
  by a factor of $\sim 100$ between Arp 220 and IRAS 15250+3609, and by
  $\sim300$ from Arp 220 to VV 114. The vertical solid line is at the
  wavelength of the NaI ionization edge.}
\label{Fig:UV}
\end{figure}

Whether the galaxies studied by Martin (2005) and by Rupke et
al. (2005a,2005b) actually have large object-to-object variations in
their near-UV fluxes is not known. We note that the results of
Goldader et al. (2002) show that near IR or even optical flux
measurements are not a good proxy for the UV flux; see the data for
VV114, IRAS 15250+3609, and Mrk 273 in Figure \ref{Fig:UV}. This
demonstrates that optical observations will not be sufficient to
answer this question. Observations by GALEX or by the ACS on the
Hubble Space Telescope could answer this question definitively.

\acknowledgments

N.M. is supported in part by the Canada Research Chair program and by
NSERC.  E.Q. is supported in part by NSF grant AST 0206006, NASA grant
NAG5-12043. Both E.Q. and C. M. were supported by grants from the
Alfred P. Sloan Foundation, and the David and Lucile Packard
Foundation.  This research has made use of the NASA/IPAC Extragalactic
Database (NED) which is operated by the Jet Propulsion Laboratory,
California Institute of Technology, under contract with the National
Aeronautics and Space Administration.


\clearpage

\clearpage

\end{document}